\documentclass{ws-ijmpd}
\usepackage[super,compress]{cite}
\begin{document}

\markboth{Y.Terada}
{Review on White Dwarf pulsars}

%
\catchline{}{}{}{}{}
%

\title{\uppercase{Summary of the session, White Dwarf Pulsars and Rotating White Dwarf Theory}}

\author{\uppercase{Yukikatsu Terada}}

\address{Graduate School of Science and Engineering, \\
Saitama University, 255 Simo-Ohkubo, Sakura-ku, Saitama 338-8570, Japan
terada@phy.saitama-u.ac.jp}

\maketitle

\begin{history}
\received{30 May 2013}
\revised{}
\end{history}

\begin{abstract}
The origin of cosmic rays remains a mystery, even over 100 years 
since their discovery. 
Neutron stars (NSs) are considered textbook cases of 
particle acceleration sites in our Galaxy, 
but many unresolved numerical problems remain. 
Searches for new acceleration sites are crucial for astrophysics. 
The magnetized white dwarfs (MWDs) have 
the same kind of rotating magnetosphere as NSs, 
and may be the source of up to 10\% of galactic cosmic ray electrons. 
In the parallel session 
of the ``white dwarf pulsars and rotating white dwarf theory'', 
we focus on the current observational results on white dwarf pulsars,
related theories of the radiation process both in white dwarfs
and neutron stars,
and the origin and rule of white dwarf pulsars,
as well as surveying on the current theories of the internal structure
and the equation of state of white dwarfs.
\end{abstract}


\ccode{PACS numbers:}


\section{Review on white dwarf pulsars}
\label{sec:intro}

\subsection{Contribution of white dwarf pulsars on galactic cosmic rays}
\label{sec:intro:cr}

The origin of cosmic rays has remained a problem 
in astrophysics for over 100 years since their discovery
by V.F. Hess in 1912. 
Among the various types of astronomical particle acceleration 
sites, rotating neutron stars (NSs) are considered to 
be a principal origin of galactic cosmic rays.
NSs can generate induced electric potentials of 
$10^{16}$--$10^{18}$~V, 
which can accelerate particles in the system, 
where the magnetic field strength $B$ is typically $10^{12}$~G, 
size $L$ is 10~km, and the spin period $P$ is on the order of
milliseconds to seconds. 
A simple analogy is a dynamo on a bicycle, 
which generates a few volts with $B \sim$ 100--1000~G, 
$L $ of a few centimeters, and $P \sim 1$~s. 
Magnetic white dwarfs (WDs) have $B \sim 10^{5}$--$10^{7}$ G, 
$L \sim 10^7$~cm, and $P \sim 1$~h 
and should have induced electric potentials of $10^{14}$--$10^{16}$~V, 
which is sufficient for particle acceleration. 
Thus, WDs can generate galactic cosmic rays. 
We call WDs having such features ``WD pulsars.'' 
Since WDs are thought to account for up to one-third of 
all astronomical objects in the Milky Way \cite{Allen73} 
and about 16\% of them have strong magnetic fields\cite{Kawka04}, 
WD pulsars should make a significant contribution to galactic cosmic rays.

The idea of the WD pulsars is not new. 
To the authors' knowledge, the first proposal of WD pulsars is found 
in a paper presented by J.P. Ostriker\cite{Ostriker68,Ostriker69},
at almost the same year that pulsar theory was proposed 
by Goldreich and Julian\cite{Goldreich69}. 
Ostriker mainly discusses the contribution of 
NS pulsars to galactic cosmic ray protons, and mentions that 
WDs can produce $10^{11}$--$10^{14}$~eV particles 
by the same acceleration mechanisms as NS pulsars. 
The first numerical estimations of the contribution of WD pulsars 
to cosmic rays were performed in 1971 by R. Cowsik\cite{Cowsik71a,Cowsik71b}.

\subsection{Observational approaches in the 1980s and 1990s}
\label{sec:intro:1980}

Magnetized WDs were already plotted in the Hillas diagram\cite{Hillas84}, 
which is often used for discussing acceleration sites in the universe. 
In 1980s and 1990s, systematic surveys of non-thermal emissions from WDs 
were performed in the radio band. 
Currently, at least seven WDs show non-thermal incoherent radio emissions 
\cite{Nelson88,Beasley94,Pavelin94,Bond02,Mason07}.
In addition, several objects such as AE Aquarii\cite{Bastian88,Abada93}
and AM Herculis\cite{Chanmugam82,Dulk83}, 
show signs of coherent radio emissions, suggesting 1--100~MeV electrons. 

In these decades, similar surveys were performed 
in the TeV gamma ray bands; 
AE Aquarii was listed as a TeV source in the first TeV catalog, 
but currently is not. 
The first detection of TeV gamma rays from AE Aquarii was 
reported by Meintjes et al.\cite{Meintjes92,Meintjes94}, 
and follow-up results in the TeV observations have been independently 
reported by several authors\cite{Bowden92,Chadwick95}, 
though detections were not confirmed by long observations 
using the Whipple telescope\cite{Lang98}. 
There are many difficulties in detecting TeV gamma rays 
using a single-dish non-imaging system, 
so the existence of WD pulsars was still an open question in the 1990s. 

\subsection{Theoretical predictions in the 1990s--2000s}
\label{sec:intro:1990}
 In the 1990s, the magnetic white dwarf AE Aquarii was interpreted as a 
``millisecond pulsar'' equivalent\cite{deJager94a}, 
in that it is a magnetic WD binary system with a high spin-down luminosity at $L_{\rm sd} = 6 \times 10^{33}$~erg~s$^{-1}$\cite{deJager94b}. 
The spin period of the object is $P = 33.0767$~s\cite{Patterson79}, 
which is the second fastest value among magnetic cataclysmic variables (CVs). 
The spin-down rate has been quite stable for more than 27 years, 
at $5.7 \times 10^{-14}$~s~s$^{-1}$\cite{Mauche06}. 
Accretion of material from the companion star onto the WD 
is expected to be low, inhibited by the propeller effect and thus generating no disc\cite{Wynn97}.  
One interpretation of the huge spin-down luminosity 
in the AE Aquarii system is the high magnetic field strength 
of $B = 50$~MG\cite{Ikhsanov98,Ikhsanov99}.
This field strength has been confirmed by the Doppler H$_\alpha$ tomogram
\cite{Ikhsanov04}. 
Using these parameters, X-ray and gamma ray emissions from the object 
have been estimated by the same group\cite{Ikhsanov06}. 
In summary, theoretical predictions of the 1990s and 2000s were 
based on standard NS-pulsar theories with magnetic reconnection. 

In the early 2010s, Terada and Dotani\cite{Terada11}
pointed out that magnetic WDs enter the magnetars' area 
in a scatter plot of spin period against magnetic momentum. 
Independently, Ikhsanov \& Beskrovnaya\cite{Ikhsanov12} pointed out that 
AE Aquarii could be a new class of magnetic CV. 
Malheiro et al.\cite{Malheiro12} suggested the possibility that 
soft gamma ray repeaters (SGRs) and anomalous X-ray pulsars (AXPs)
could be interpreted not as magnetars 
but as magnetic WDs, 
although this cannot explain the big problem of SGRs and AXPs, 
that their X-ray flux exceeds dipole luminosities. 
Note that efficiency from the dipole luminosity into the X-ray luminosity of one magnetic WD, EUVE J0317-855, was recently measured to be very low as compared with magnetars\cite{Harayama13}.

\subsection{Surveys for non-thermal emissions after 2005}
\label{sec:intro:2000}

Since the latter half of the 2000s, several observational searches for 
possible WD pulsars have been performed 
in the radio, X-ray, and gamma ray bands. 
In 2005, Zhang and Gil\cite{Zhang05} reported a transient radio source 
in the galactic center region, which they interpreted as a transient WD pulsar. 
In the X-ray band, the X-ray satellite Suzaku\cite{Mitsuda07} 
searches for possible non-thermal hard X-ray emissions from a WD, 
which should be radiated from accelerated particles in the system. 
In 2007, Terada et al.\cite{Terada08} reported discovery of hard X-ray 
pulsation from a WD binary system, AE Aquarii, 
claiming that this sharp pulsation was similar to those in NS pulsars 
and would have a non-thermal origin from analysis of the X-ray spectra.
In TeV bands, the gamma ray telescope MAGIC observed AE Aquarii 
in coordinated observations between X-ray and radio bands, 
but no significant detection in the TeV band has been 
reported\cite{Sidro08,Mauche12}. 
Another gamma ray telescope, H.E.S.S., also independently performed 
a coordinated observation of AE Aquarii in X-ray and optical bands. 
The details are summarized in section \ref{sec:suzaku:multi}.

\subsection{Session Overview}
\label{sec:intro:so04}
On the parallel session, the current results and plan of X-ray observations are presented by Y.Terada, T.Hayashi, A.Harayama, and T.Kitaguchi, which is summarized in section 2. From the theoretical approach, T. Wada has shown the results on particle simulations on the WD magnetosphere (session \ref{sec:theory:wdns}). There are presentations and discussion on the origin of the class of these objects by M. Malheiro, C. J. Goulart, N. Ikhsanov,  K.Kashiyama, and other twoWDs than WD pulsars are introduced by S. Mereghetti and A. Nucita (session \ref{sec:theory:wdorigin}). The general relativity and the internal structure of WDs is also discussed (session \ref{sec:theory:eos}).

\section{Observational Approaches on the Study of White Dwarf pulsars}
\label{sec:obs}

\subsection{Summary of the X-ray observations}
\label{sec:obs:summary}
Normally, magnetic CVs emit strong thermal bremsstrahlung 
from hot plasmas on the magnetic poles of WDs\cite{Patterson94,Terada01,Terada04}. 
In searches for possible signals from high energy particles 
in magnetic CV systems, it is important to avoid such strong 
thermal emission. 
In harder X-ray bands, thermal emissions become dimmer.
Suzaku is highly sensitive in the hard X-ray band\cite{Mitsuda07}, 
and as mentioned in section \ref{sec:intro:2000} 
Terada et al.\cite{Terada08} reported the discovery of spiky pulsations 
in the 4--25~keV band from AE Aquarii using the Suzaku satellite. 
Note that they do not conclude that this pulsation is non-thermal, 
because of limitations in the statistics and sensitivities of the data, 
they just point out the possibility of non-thermal origins of this emission.

In the parallel session, three authors (Y.Terada, T.Hayashi, and A. Harayama)
show recent results on X-ray surveys with the X-ray satellite, Suzaku,
as summarized in session \label{sec:suzaku:summary}.
The very recent mission, NuSTAR, has quite high sensitivities in the
hard X-ray band, and thus it should be the very powerful tool for
studies on the particle acceleration process of white dwarf pulsars.
T. Kitaguchi summarized the latest status of NuSTAR mission
and showed the future plan of the observation of WD pulsars.

\subsection{Summary of the X-ray observations with Suzaku}
\label{sec:obs:suzaku}
After the discovery of hints of high energy particles in the AE Aquarii system\cite{Terada08}, 
Y.Terada et~al set up a Suzaku-based project for systematic searches for non-thermal emissions from WDs. 
The targets are not only binary systems (magnetic CVs), 
but also isolated WDs. 
Table \ref{tbl} summarizes the observed objects. 

\begin{table}
\caption{Summary of Suzaku observations}
\label{tbl}
\begin{center}
\begin{tabular}{llcccc}
\hline
Object Name & Type & Year & Exposure & OBSID & Ref \\
\hline
AE Aqr & binary& 2005 &  70 ks & 400001010 & Terada+ 08\cite{Terada08}\\
AE Aqr & binary& 2006 &  50 ks & 400001020 & Terada+ 08\cite{Terada08}\\
AM Her & binary& 2008 & 100 ks & 403007010 & Terada+ 10\cite{Terada10}\\
AE Aqr & binary& 2009 & 160 ks & 404001010 & section \ref{sec:suzaku:multi}\\
EUVE J0317-85.5 & isolated & 2009 & 60 ks & 404019010 & Harayama+ 13\cite{Harayama13}\\
IGRJ00234+6141 & binary&  2010 & 80 ks & 405022010& (prep)\\
V2487 Oph & binary& 2010 & 50 ks & 405021010 & (prep)\\
EUVE J1439+75.0 & isolated & 2012 & 40 ks & 407039010 & \\
PG 1658+440 & isolated & 2012 & 50 ks & 407040010 & \\
\hline
\end{tabular}
\end{center}
\end{table}

In their surveys of binary WDs, 
they implemented the following three strategies:
\begin{itemize}
\item To confirm that AE Aquarii is a WD pulsar, 
Terada et~al performed a coordinated observation of the object with optical 
telescopes, the Suzaku X-ray satellite, and the H.E.S.S. gamma ray telescope. 
The results are summarized in  section \ref{sec:suzaku:multi}.

\item To avoid thermal emission in the soft X-ray band, 
they used Suzaku to observe another TeV candidate, AM Herculis, 
in a very low state at which mass accretion from the companion star is suppressed.
As a result, Terada et~al found a possible non-thermal component in quiescent phases, although it is statistically difficult to distinguish from a multi-temperature plasma model\cite{Terada10}.

\item To efficiently select candidates emitting hard X-ray non-thermal radiation, 
Terada et~al used the hard X-ray catalog with Swift\cite{Cusumano10} and INTEGRAL\cite{Landi08,Bird10,Scaringi10} to pick up objects 
with hard X-ray spectra; that is, temperatures are reported 
to be very high compared with other magnetic CVs. 
Such hard objects may contain both hard non-thermal emissions 
and normal thermal bremsstrahlung. 
Among 39 objects, they selected the top two hardest magnetic CVs, 
which are V2487 Oph and IGR J00234+6141 (Table \ref{tbl}). 
As summarized by T. Hayashi, they found that one of these two objects,
V2487 Oph, requires a power-law emission in the X-ray spectrum,
indicating that it is another white dwarf pulsar.
\end{itemize}

In a survey of isolated WDs, 
Terada et~al first picked up 480 magnetized WDs\cite{Kawka07} 
with magnetic field strength $10^4$--$10^9$ G,
from a WD catalog\cite{Eisenstein06} by the Sloan Digital Sky Survey (SDSS) 
containing about 9000 objects.
Among them are 82 objects for which both $P$ and $B$ are known. 
The white dwarf EUVE J0317-855 has the largest magnetic dipole energy, 
and they thus targeted it for observation using Suzaku. 
A. Harayama reported the results on this object\cite{Harayama13},
but no X-rays have been detected from this object.
They continue the survey for possible non-thermal emission from isolated WDs;
they checked the ROSAT archive data of 247 magnetic WDs, 
whose magnetic field strengths are larger than $10^6$~G, 
and found three objects, EUVE J1439+75, PG 1658+440, and EUVE J0823-25.4,
that show soft X-ray emissions. 
They therefore performed Suzaku-based deep hard X-ray observations of 
the first two WDs listed in Table \ref{tbl}. 
They will report the results in a future paper.

\subsection{Multi-wave campaign of AE Aquarii in Optical, X-ray, and TeV gamma-ray}
\label{sec:suzaku:multi}

\begin{figure}[hb]
\centerline{\psfig{file=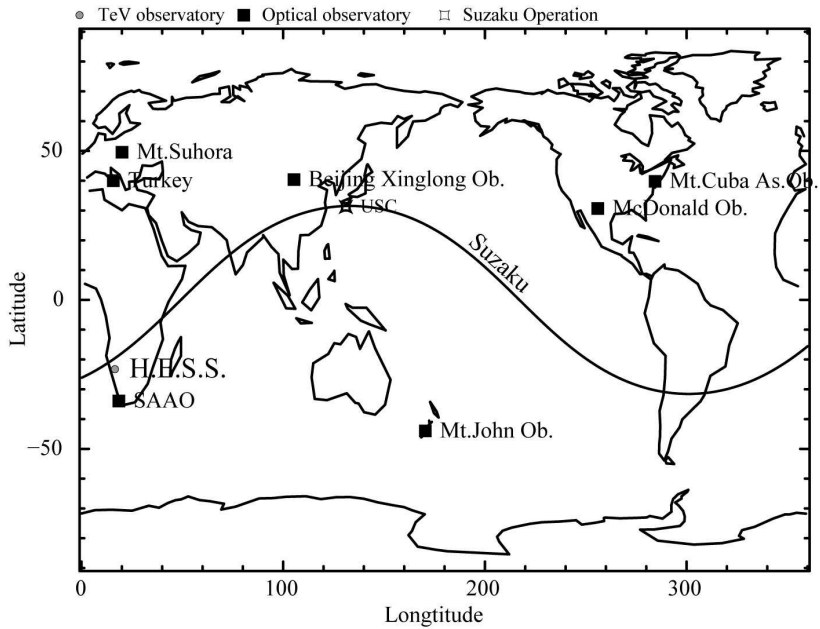,width=0.85\textwidth}}
\vspace*{8pt}
\caption{Observatories in the simultaneous campaign of AE Aquarii in 2009. \label{f1}}
\end{figure}

This is reported by Y. Terada.

The physical condition of AE Aquarii may be somewhat different 
from those of NS pulsars; 
the region up to the light cylinder of a magnetized WD is 
filled with accreting materials from the companion, 
with a density probably much higher than 
the Goldreich-Julian critical value\cite{Goldreich69}.
This would reduce the effective electric potential for accelerating particles. 
One possible solution is that the acceleration site is rather close to the WD, 
where the density becomes very low due to the propeller effect. 
To elucidate this scenario, 
measurement of the magnetic field strength of the non-thermal emission 
region is crucial. 
While the non-thermal X-ray emission from AE Aquarii is likely to be 
of synchrotron origin, TeV emission must be due to the inverse Compton process, 
and thus they can estimate the magnetic field strength of the emission 
from X-ray and TeV gamma ray measurements. 
Therefore, they performed a simultaneous observation of AE Aquarii 
in the optical, X-ray, and TeV gamma ray bands in October 2009. 
It was the first observation of the TeV telescope, H.E.S.S., 
in the guest observation program of this telescope. 
The TeV observation was reserved for two hours of each of the four nights from 16 to 19 October. 
The X-ray observation with Suzaku was scheduled 
from 20:00 UT 16 October to 5:26 UT 20 October, 
with an effective exposure of 160~ks. 
The optical observation was covered by seven observatories to cover all latitudes (Fig.~1). 
The time allocated for AE Aquarii was from 14 to 22 October.

Unfortunately the weather was too poor to detect any signals from AE Aquarii 
with H.E.S.S., and thus they cannot judge whether non-detection
of TeV gamma rays was due to sensitivity limitations or 
the nature of the object.
The situation was similar for the optical telescopes, 
but they got several light curves during the campaign with the NAOO, SAAO, and TURK observatories. 
The X-ray observation with Suzaku was performed over 3.5 days.
In the XIS data, they found more than 10 flares, 
which are clearly seen in the soft energy band (0.5--2 kev), 
as already reported by Terada et al.\cite{Terada08}.
Among them, one flare at MJD 55122.839 was detected in both 
the two optical telescopes and the Suzaku satellite (Fig.~2). 
\begin{figure}[hpb]
\centerline{\psfig{file=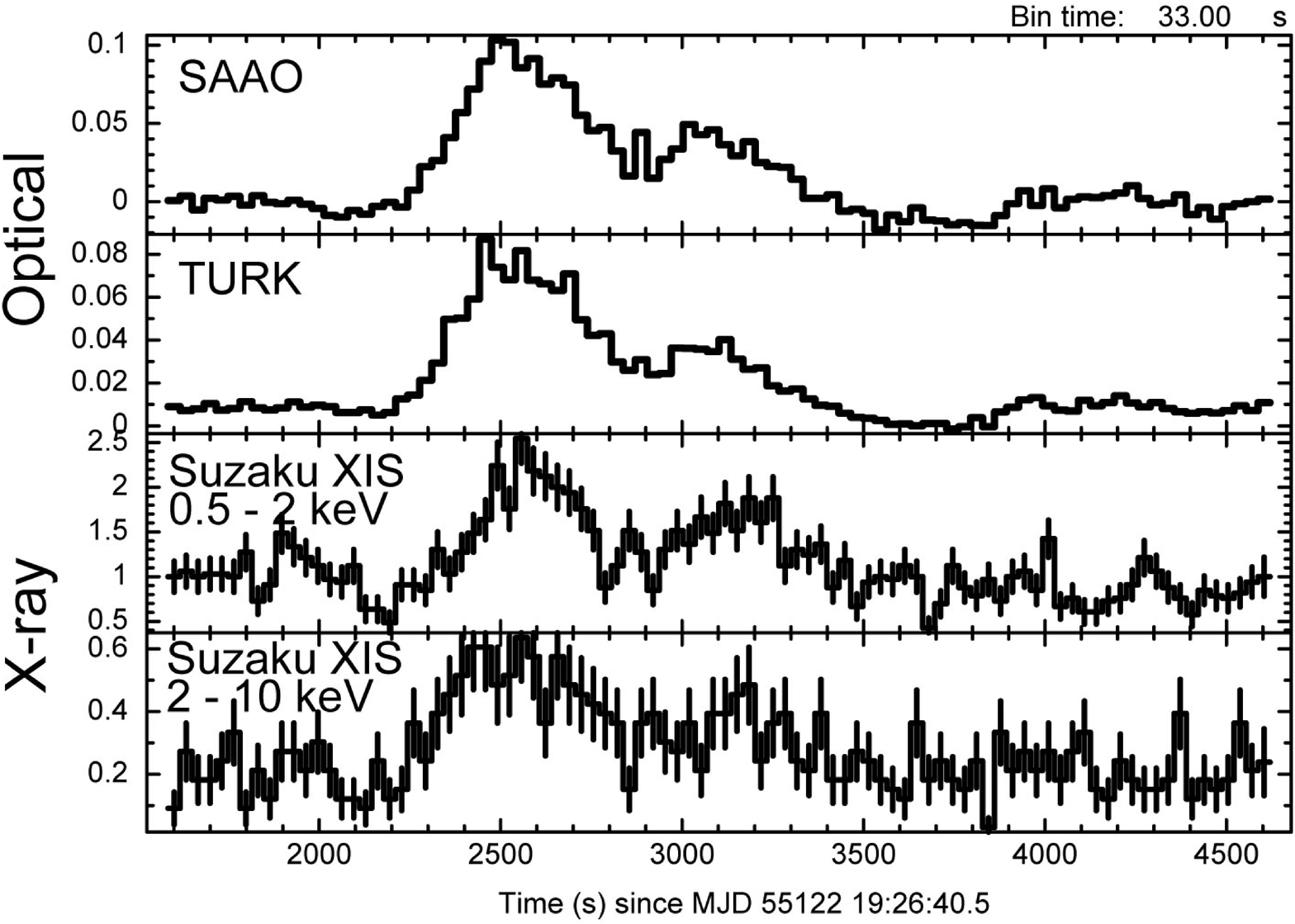,width=0.7\textwidth}}
\vspace*{8pt}
\caption{Flare events shown in optical telescopes, SAAO and TURK, and X-ray bands. \label{f2}}
\end{figure}

\begin{figure*}[htpb]
\centerline{
\psfig{file=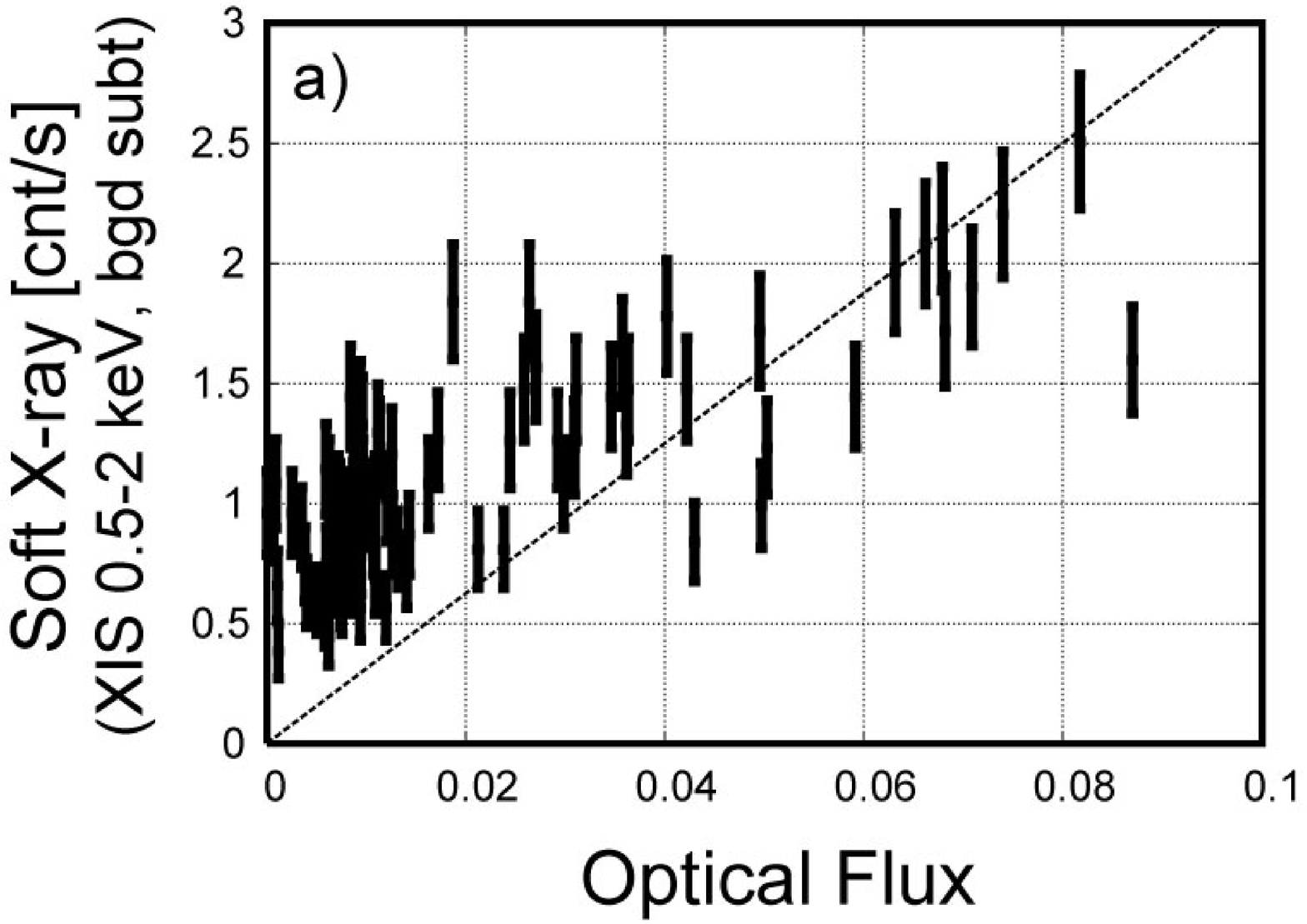,width=0.4\textwidth}
\psfig{file=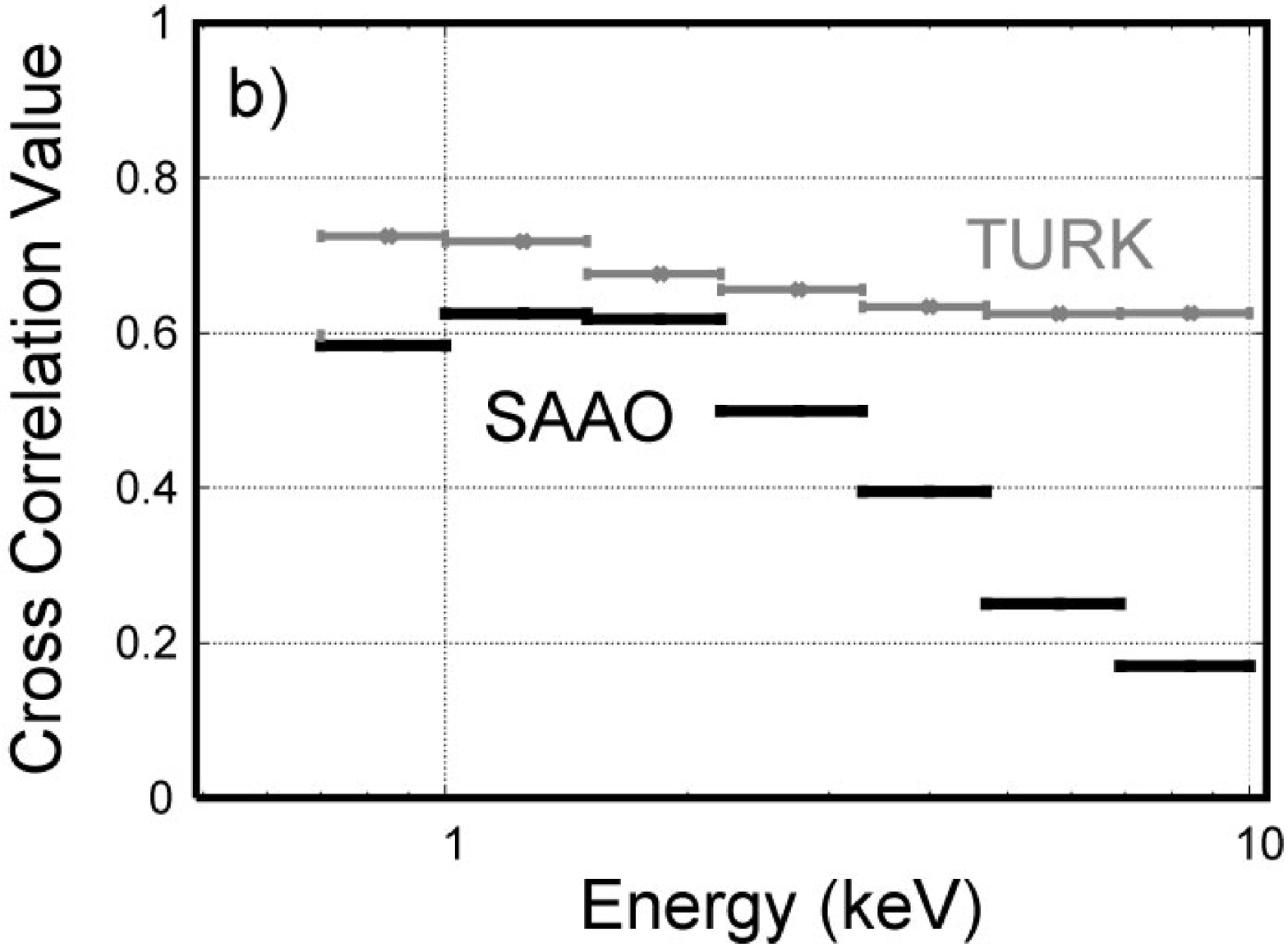,width=0.4\textwidth}
}
\vspace*{8pt}
\caption{(left) Scatter plot between the count rates in the optical band and those in the soft X-ray band. (right) Cross-correlation between light curves in the X-ray and optical bands. \label{f3}}
\end{figure*}

The origin of the soft X-rays is mainly explained by 
accretion powered emission (thermal electron bremsstrahlung)\cite{Terada08},
although rotation-powered emission (or non-thermal emission, here) is 
contaminated by a small fraction.
As demonstrated in Figure 3 (left), 
the optical flare should be derived by mass accretion 
because the behavior in the optical light curve is similar 
to that in the soft X-rays. 
If they assume that the optical emission is purely accretion-powered emission
(there is no contamination of non-thermal radiation), 
they can check how much of the non-thermal emission is accounted for 
in the X-ray photons. 
Figure 3 (right) shows a cross-correlation between the X-rays and optical
light curves during the flare,
representing less contribution of thermal radiation 
in the harder X-ray band.
Therefore, in the above 2~keV band, non-thermal emission should exist.
They will report details in a future paper.

\begin{figure}[htpb]
\centerline{\psfig{file=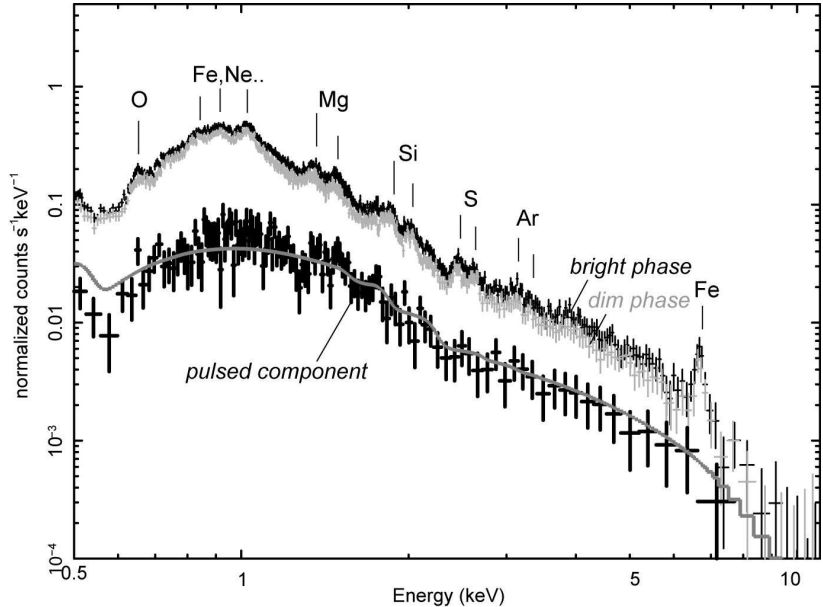,width=0.85\textwidth}}
\vspace*{8pt}
\caption{Phase resolved X-ray spectra of AE Aquarii with Suzaku. Black thin and dim crosses represents spectra in peak and bottom spin phases, and thick black crosses show the spectrum in a pulsed component, i.e., spectrum in bright subtracted by that in dim phases. Thin lines over the thick crosses is the best-fit power-law model of the pulsed component.\label{f4}}
\end{figure}

Combining the 2005, 2006, and 2009 datasets of observations 
of AE Aquarii with Suzaku, they derived the spin period $P$ and the
period derivative $\dot{P}$ as $P = 33.07742(1)$~s and 
$\dot{P} = 10^{-5.8 \pm 2.5}$~s/year. 
These values are consistent with previous works\cite{deJager94a,Mauche06}.
Then, they performed the spin-phase resolved analyses of the X-ray spectra,
combining all three epochs (Fig.~ 4).
In the bright and dim spin phases, they can see many atomic lines 
in the spectra, indicating that the emission comes from the thermal plasma, and so the origin is mainly thermal.
However, these atomic lines disappear completely in the spectra in the pulsed component. 
Without the third long observation in 2009, 
this result is suggested but statistically insufficient.
Numerically, if they reproduce the spectrum of the pulsed component
by a thermal model (MEKAL), then the temperature is $1.65\pm0.16$~keV and 
the metal abundance becomes very low as $0.08_{-0.04}^{+0.06}$~solar 
($\chi^2$/d.o.f = 0.95),
statistically suggesting that the emission is non-thermal.
The pulsed emission is well reproduced by the power-law model 
with the photon index of $2.28\pm 0.08$ 
with the X-ray luminosity of $1.8 \times 10^ {30}$~erg/s 
in the 0.5 to 10~keV band, 
which corresponds to 0.03\% of the spin down luminosity. 
They will report more detailed results in another paper in the near future.

In this section \ref{sec:suzaku:multi}, 
they thank the following people who contributed to the simultaneous observation of AE Aquarii: A. Bamba (Aoyama Gakuin, Japan),
K. Makishima (U.Tokyo, Japan), 
T. Hayashi, A. Harayama T. Dogtani, M. Ishida (ISAS/JAXA, Japan), and
K. Mukai (NASA/GSFC, USA) 
for continuous support in X-ray observations;
Y. Urata (Taiwan Cent Univ, Taiwan) and A. Nitta (Gemini Observatory, USA) 
for optical observations; and
Okkie de Jager (North-West Univ, South Africa), Paula Chadwick (Durham Univ, UK), Stefan Wagner (LSW, Germany), and David Buckley (Univ. of Cape Town, South Africa)
for TeV gamma ray observations.
This work in this section 
was supported by a Grant-in-Aid for Scientific Research (B) 
from the Ministry of Education, Culture, Sports, Science and 
Technology (MEXT) (No. 23340055).

\section{Theoretical approaches on WD pulsars}
\label{sec:theory}

\subsection{Similarities between NS and WDs}
\label{sec:theory:wdns}
T.Wada et al has shown the current results from the particle simulation for rotating white dwarf magnetosphere, using the simulation code which originally developed for neutron stars. They show a steady solution with accelerating regions around the star. In the discussion, they also provided the death line for WD pulsars.

The similarity between magnetized WD and SGRs\&AXPs (soft gamma-ray repeaters and anomalous X-ray pulsars) was proposed by M. Malheiro and C. J. Goulart\cite{Malheiro12}. This idea is originally triggered by the diagram of NSs and WDs between the period and the magnetic momentum in Terada and Dotani 2010\cite{Terada11}. On this idea, if we interpret the SGRs\&AXPs as the magnetized WD, not the magnetars, the emission region becomes larger because the radius of the object is larger than NS cases, and then we do not need the stronger surface-magnetic-field strength than the critical value. After long discussions in the parallel session, we conclude the followings.
\begin{itemize}
\item the idea may work as one of the interpretations of SGRs\&AXPs
\item the idea cannot explain the fundamental problem on this class of object that the high energy emission from SGRs\&AXPs exceeds the dipole energy of the object, i.e., the origin of the energy source of the radiation
\end{itemize}

\subsection{Where the WD pulsars comes and what affects on}
\label{sec:theory:wdorigin}

On the history of the WD pulsars, N. Ikhsanov discussed the formation of the white dwarf pulsar, taking AE Aqr as an example; i.e., why such highly spinning down object exist. 
On the rules of WD pulsars, K.Kashiyama proposed the possibility of WD pulsars as the e+ e- sources in our Galaxy\cite{Kashiyama11}. He calculated the contribution of WD pulsars on the TeV e+e- flux observed by PAMELA\cite{Adriani11} numerically and conclude that the total energy budget is comparable with NSs and pointed out that WD pulsars can accelerate e+ e- into TeV energies with longer time scale.

In the parallel session, there are presentations on other WDs, RX J0648.0-4418 by S. Mereghetti \cite{Mereghetti09}and 1RXS J180431.1-273932 by A. Nucita.

\section{General Relativisty and Internal Structure of White dwarfs}
\label{sec:theory:eos}

White dwarfs are one of the best laboratories to test the fundamental physics of the Fermi gas under an extreme condition. In the simple model by Chandrasekhar \cite{Chandrasekhar35}, the macroscopic parameters, mass and radius, of WDs are defined by the balance between the degenerated pressure of electrons and the gravitational force mainly by protons. In the parallel session, four authors have shown the effects of physical parameters on the equation of state (EoS) of WDs. 

K. Boshkayev has shown the effects of the spin rotation under general relativistic case \cite{Rotondo11}. They also showed that the minimum rotation periods are approximately 0.3, 0.5, 0.7 and 2.2 seconds for a rotating $^4$He, $^{12}$C, $^{16}$O, and $^{56}$Fe WDs. S. Filippi has shown the effect on the analog gravity for non rotating WDs, in which the electrons are not perfect fluid but behave under hydrodynamical equations with the effective geometry formalism. U. Das has shown the effects of the magnetic field on the EoS of WD numerically and got a result that the mass and radius relation has turning points which correspond to the kinks in the EoS and that the maximum mass became 2.3 M$_\odot$, i.e., a super-Chandrasekhar WDs. Finally, S. Martins has shown the effects of temperature on EoS of WD.


\end{document}